\documentclass[twocolumn]{openjournal}
\usepackage{natbib}
\usepackage{graphicx}
\usepackage{graphicx,amsmath,amssymb,amstext}
\usepackage{amsbsy,amsfonts,amsthm,color}
\usepackage{caption}
\usepackage{subcaption}
\usepackage{xcolor}
\usepackage{graphicx}
  \usepackage{lipsum}
\usepackage[colorlinks,linkcolor=blue,citecolor=blue,urlcolor=blue ]{hyperref}

\usepackage{soul}

\usepackage{xcolor}

\newcommand\blfootnote[1]{%
  \begingroup
  \renewcommand\thefootnote{}\footnote{#1}%
  \addtocounter{footnote}{-1}%
  \endgroup
}

\begin{document}

\title{\textbf{FRBSTATS: A web-based platform for visualization of\\fast radio burst properties}}

\author{Apostolos Spanakis-Misirlis$^{1,2}$}
\author{Cameron L. Van Eck$^{3}$}

\blfootnote{E-mail: \href{mailto:0xcoto@protonmail.com}{0xcoto@protonmail.com}}

\affiliation{\\$^1$ Department of Informatics, University of Piraeus, Greece\\
$^2$ Institute of Astrophysics, FORTH, Dept. of Physics, University of Crete, Voutes, University Campus, GR-71003, Heraklion, Greece\\
$^3$ Research School of Astronomy and Astrophysics, The Australian National University, Canberra ACT 2611, Australia\\}

\begin{abstract}
The study of fast radio bursts (FRBs) is of great importance, and is a topic that has been extensively researched, particularly in recent years. While the extreme nature of FRBs can serve as a tool for researchers to probe the intergalactic medium and study exotic aspects of the Universe, keeping track of FRB properties is challenged by the frequent detection of new bursts. We introduce the FRBSTATS platform, which provides an up-to-date and user-friendly web interface to an open-access catalogue of published FRBs, along with a statistical overview of the observed events. The platform supports the retrieval of fundamental FRB data either directly through the FRBSTATS API, or in the form of a CSV/JSON-parsed database, while enabling the plotting of parameters and their distributions, for a variety of visualizations. These features allow researchers to conduct population studies and comparisons with astrophysical models, describing the origin and emission mechanism behind these sources. So far, the inferred redshift estimates of 813 bursts have been computed, providing the first public database that includes redshift estimates inferred from dispersion measure entries for nearly all observed FRBs, as well as host redshifts (where available). Lastly, the platform provides a visualization tool that illustrates associations between primary bursts and repeaters, complementing basic repeater information provided by the Transient Name Server. In this work, we present the structure of the platform, the established version control system, as well as the strategy for keeping such an open database up to date. Additionally, we introduce a novel, computationally-efficient, clustering-based approach that enables unsupervised classification of hundreds of bursts into repeaters and non-repeaters, resulting in the discovery of one new FRB repeater. The platform is accessible at: \url{https://www.herta-experiment.org/frbstats}
\end{abstract}

\maketitle

\section{Introduction}
\label{sec:introduction}

\begin{figure*}
	\includegraphics[width=\textwidth]{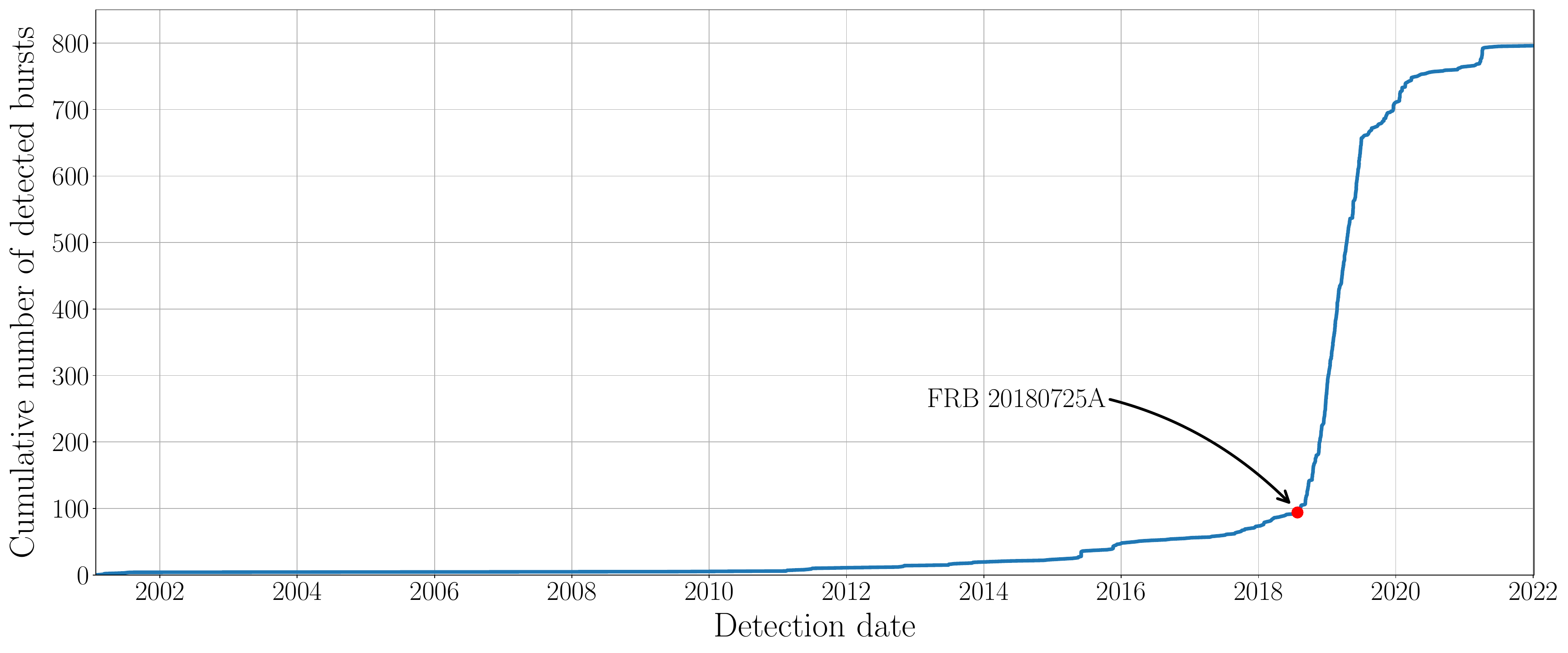}
    \caption{Total number of published bursts (from 622 unique FRB sources) observed as a function of their detection time. The notable increase of observed bursts in July 2019 came with the commissioning of the Canadian Hydrogen Intensity Mapping Experiment Fast Radio Burst (CHIME/FRB) Project. FRB 20180725A marks the very first detection by CHIME, representing the first ever FRB observed below 700 MHz.\\}
    \label{fig:count}
\end{figure*}

Fast radio bursts (FRBs) are bright, short-duration pulses of unknown origin, observed at radio wavelengths. Since the first detection by \citet{lorimer2007}, hundreds of such events have been observed, with varying emission characteristics (see Fig.~\ref{fig:count}). Despite the 15-year period of FRB observations however, the emission mechanism behind these sources remains unknown \citep{petroff2022}. Although the exact origin of FRBs is highly debated, nearly all FRBs detected so far are extragalactic, with varying spectral, polarization, and time-domain emission properties. With the exception of the Galactic magnetar SGR 1935+2154 \citep{andersen2020}, the dispersion measure of FRBs implies that their origin is extragalactic. It is worth noting, that some FRBs exhibit a repeating behaviour \citep{spitler2016}, which has ruled out, altered, and even introduced a variety of new theoretical models describing the radiation process \citep{platts2019}. An unusual subclass of repeating FRBs appears to show a periodic activity \citep{amiri2020,rajwade2020}, which may further narrow down the list of candidate models.

Because of their enigmatic nature and the rapid rate of discoveries, gaining a better and more complete understanding of these astrophysical objects is an essential prerequisite to uncovering a significant piece of the puzzle behind the mysterious presence of this class of sources. Due to the unpredictability of FRBs however, the consistent monitoring of individual sources is a highly time-consuming and challenging task. While each observation comes with its own derived properties, carrying out literature reviews for each individual burst becomes impractical, especially as the number of detected bursts increases. Without the most significant properties of each detected burst indexed in an accessible database, population studies and extensive analysis of FRBs (taking every observed event into consideration to build statistical models) become laborious.

While previous efforts relevant to cataloging FRBs, such as the Transient Name Server \citep[TNS;][]{yaron2020} and FRBCAT \citep{petroff2016} have been able to collect a meaningful amount of data from FRB publications, such platforms have struggled to remain up to date with new bursts (occasionally with several values being reported inaccurately), and are arguably limited in terms of their functionality and ease of use.

In this work, we present a platform that hosts an accurate and complete catalogue of FRBs, with the primary goal of enabling researchers to easily derive fundamental properties of bursts, with minimal effort. In Sect.~\ref{sec:architecture}, we begin by describing the high-level architecture of the system, as well as the strategy for maintaining and keeping the catalogue up to date, as autonomously as possible. Accessible via a user-friendly web interface, FRBSTATS includes an open-access catalogue of all FRBs published to date (Sect.~\ref{sec:catalogue}), along with a statistical overview of their properties. Furthermore, the platform hosts a page that illustrates the association between primary bursts and repeaters (Subsect.~\ref{sec:repeaters}), complementing basic repeater information provided by the TNS.\footnote{\url{https://www.wis-tns.org}} In addition to the ability of plotting of parameter distributions for a variety of visualizations (Subsect.~\ref{sec:plots}), the platform supports the retrieval of burst data, directly through the FRBSTATS API (Subsect.~\ref{sec:api}). We finally conclude by summarizing our work and discussing future enhancements for the platform (Sect.~\ref{sec:conclusions}).\\

\section{Platform architecture}
\label{sec:architecture}

\begin{figure}
	\includegraphics[width=\columnwidth]{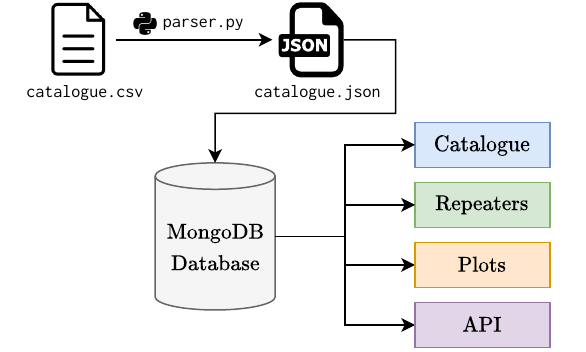}
    \caption{System architecture of the FRBSTATS platform, indicating the flow of data from one subsystem to another. The \texttt{parser.py} process refers to a Python script that is part of the CI/CD pipeline. Its role is to convert the input CSV catalogue to a JSON-formatted file for improved accessibility by the MongoDB database infrastructure.}
    \label{fig:architecture}
\end{figure}

In order to ensure all data and code supporting the platform is findable, accessible, interoperable, and reusable, the source of the website is hosted on a Git repository.\footnote{\url{https://github.com/HeRTA/FRBSTATS}} This enables and encourages researchers to openly contribute with code, data, and ideas for improvement. Additionally, the project utilizes continuous integration and continuous deployment (CI/CD) pipelines, which address the problem of delayed development and updates to the platform. Fig.~\ref{fig:architecture} describes the high-level architecture of the FRBSTATS platform.

A spreadsheet is available (link provided in the repository along with contribution guidelines), where users can report new bursts to be picked up by the database. The properties of each burst are automatically distributed to the corresponding subprocesses for further management, and any new burst reported on the TNS is automatically queried and merged to FRBSTATS, on a daily basis.

Based on a non-relational document model, MongoDB\footnote{\url{https://github.com/mongodb/mongo}} offers a suitable NoSQL database, which serves as the catalogue infrastructure for the platform. By automatically parsing the CSV catalogue (obtained from the spreadsheet) as a JSON-formatted file, the catalogue page permits quick search queries, based on either the designation of the burst of interest, or its parameters.

\section{Catalogue}
\label{sec:catalogue}

The catalogue consists of the observed properties for each burst: the topocentric Coordinated Universal Time (UTC) and Modified Julian Date (MJD) of the observed event, the corresponding telescope that detected the burst, and its sky location in equatorial (right ascension $\alpha$ and declination~$\delta$, along with their associated errors), as well as Galactic coordinates (longitude $l$ and latitude~$b$). Additionally, the database includes the frequency of each burst, its dispersion measure and error, host and redshift estimate inferred from the dispersion measure \citep[computed with the help of the \texttt{fruitbat}\footnote{The package takes the coordinates and dispersion measure of the burst, and based on the YMW16 free-electron distribution model \citep{yao2017}, the tool then computes an estimate for the redshift based on the derived excess dispersion measure.} package;][]{batten2019}, peak flux density, burst width and fluence, as well as the signal-to-noise ratio of the detection. For the redshift inference, the \texttt{fruitbat} package calculates the excess dispersion measure by subtracting the contribution of the Galactic foreground, before using a relation between $\mathrm{DM}_{\mathrm{IGM}}$ and $z$ \citep{ioka2003,inoue2004,zhang2018_fruitbat}. The catalogue also contains a reference for each burst to direct users to the original data source.

Event data are obtained from the TNS, FRBCAT, the First CHIME/FRB Fast Radio Burst Catalog,\footnote{\url{https://www.chime-frb.ca/catalog}} The Astronomer's Telegram\footnote{\url{https://www.astronomerstelegram.org}} \citep{rutledge1998}, and the corresponding publications cited on the FRBSTATS catalogue. We note that since Virtual Observatory Events (VOEvents) from CHIME\footnote{\url{https://www.chime-frb.ca/voevents}} have not been verified and characterized, they have not been incorporated into the catalogue. Since the TNS remains highly relevant for the submission of new FRB observations, appropriate CI/CD pipelines have been set up and scheduled to monitor the database for new entries on an hourly basis, updating the FRBSTATS catalogue automatically. For some bursts, parameters obtained and merged from other catalogues have been reviewed and corrected based on the publications associated with each burst.

\subsection{Repeaters}
\label{sec:repeaters}

As mentioned in Sect.~\ref{sec:introduction}, a fraction of FRB sources appear to exhibit a repeating activity; a single FRB source is capable of producing more than one burst. This clue becomes especially significant if the observed properties appear very similar to the previously-emitted bursts.

The discovery of the first repeater (FRB 20121102) by \citet{spitler2016} demonstrated that such a class of FRBs could not originate from cataclysmic events involving the collapse of supramassive neutron stars \citep{falcke2014} or neutron star mergers \citep{hansen2001}. Although it is not yet clear whether a single emission mechanism is responsible for all FRBs (repeaters and non-repeaters), the repository of FRB theories\footnote{\url{https://frbtheorycat.org}} \citep{platts2019} consists of studies that rule out even more theories for the progenitor of FRB repeaters, such as pulsar-black hole mergers \citep{bhattacharyya2017}, binary white dwarf mergers \citep{kashiyama2013}, the collapse of strange-star crusts \citep{zhang2018}, neutron star-white dwarf collisions \citep{liu2018}, among other theories.

While the study of FRB repeaters is a valuable method for investigating the potential emission mechanisms responsible for each burst, the manual classification of FRBs into repeaters and non-repeaters (one-offs) can be a tedious task. FRBSTATS makes this process easier, by automatically identifying repeating bursts and distinguishing them from one-off events. This enables researchers to easily associate the observed properties of each burst, without having to investigate the parameters of hundreds of events.

\subsubsection{Repeater classification using DBSCAN}
\label{sec:dbscan}

Considering the vast number of FRBs detected to date, in order to characterise bursts into repeaters and one-offs, an automated classification method is required. We describe a novel method for automatically classifying repeaters from non-repeaters using density-based spatial clustering of applications with noise (DBSCAN). Originally proposed by \citet{ester1996}, this data clustering algorithm is ideal for certain applications that require the grouping of a set of neighbouring points.

With the objective of rapidly classifying nearly a thousand bursts into repeaters and non-repeaters, we find that this approach can be applied to FRBs using merely three parameters: the candidate's coordinates ($\alpha$, $\delta$), as well as its dispersion measure (DM).

The DM describes the integrated free-electron column density between the observer and the FRB of interest,\\ and based on the model of \citet{macquart2020}, the observed DM can be split into four contributing factors,
\begin{align}
    \mathrm{DM_{\mathrm{FRB}}}=\mathrm{DM}_{\mathrm{disk}}+\mathrm{DM}_{\mathrm{halo}}+\mathrm{DM}_{\mathrm{cosmic}}+\mathrm{DM}_{\mathrm{host}},
	\label{eq:dm}
\end{align}
where the contribution of the Galactic disk and halo, the extragalactic distribution of ionized gas and the host galaxy of the FRB source are all taken into account.

Despite the uncertainty of $\mathrm{DM_{\mathrm{host}}}$ described by \citet{james2022}, \citet{yang2017} show that variations in the dispersion measure of repeating FRBs are small. The stability of this parameter makes the DM a great alternative to using source distance as a clustering parameter, which, due to the limitation of free-electron distribution models being confined to Galactic sources \citep{cordes2004, schnitzeler2012}, remains unknown.

Assuming the three parameters ($\alpha, \delta, \mathrm{DM}$) of repeaters used for clustering possess negligible variation over time/remain constant, we can employ a custom metric to reliably evaluate the distance between two given points (FRBs), using the three-dimensional Euclidean distance that separates them.

Let $x_1, y_1, z_1$ and $x_2, y_2, z_2$ be the Cartesian coordinates of $\mathrm{FRB}_1$ and $\mathrm{FRB}_2$, respectively. The position of each burst expressed in terms of $\alpha$, $\delta$ and DM is then given by
\begin{align} \label{eq:xyz}
    x_i &= \mathrm{DM}_i \cos{\delta_i} \sin{\alpha_i} \\
    y_i &= \mathrm{DM}_i \cos{\delta_i} \cos{\alpha_i} \\
    z_i &= \mathrm{DM}_i \sin{\delta_i},
\end{align}
where $i \in \{1,2\}$ refers to the FRB of interest. We can then compute the Euclidean distance $d$ between $\mathrm{FRB}_1$ and $\mathrm{FRB}_2$ using
\begin{align*}
    d\big(\mathrm{FRB}_1, \mathrm{FRB}_2\big) = \sqrt{\big(x_1-x_2\big)^2+\big(y_1-y_2\big)^2+\big(z_1-z_2\big)^2}.
\end{align*}

\begin{figure*}
	\includegraphics[width=\textwidth]{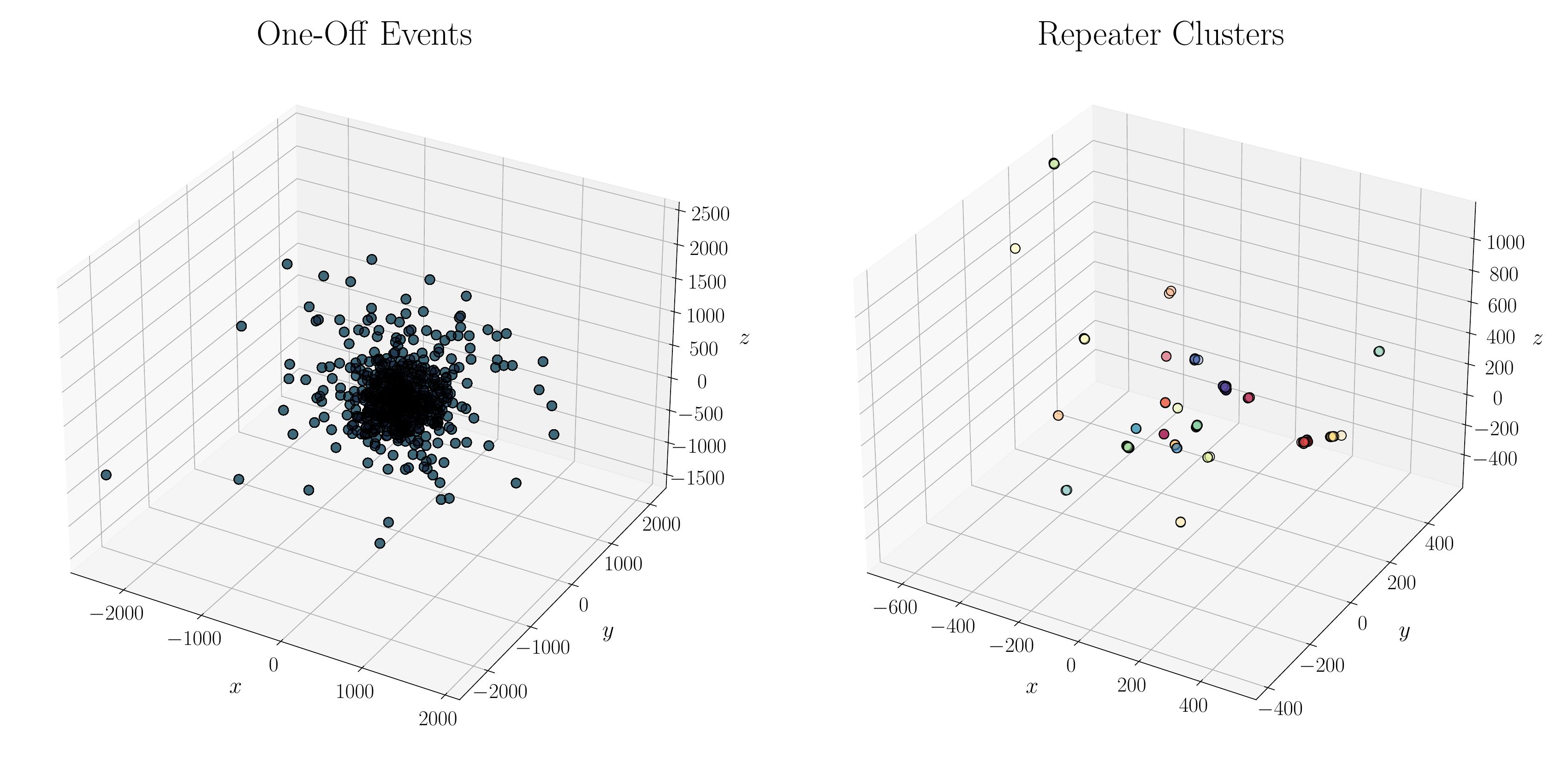}
    \caption{Output scatter plots resulting from the DBSCAN algorithm for $\epsilon=1.15$. The one-off events are counted as noisy samples/stray data points (left), while the detected 23 clusters are grouped and represented with a corresponding colour associated with an individual repeater (right).\\}
    \label{fig:repeaters}
\end{figure*}\pagebreak
However, considering the error values of $\alpha$, $\delta$ and DM are commonly reported, it is important to incorporate uncertainties into the clustering analysis so that the resulting clusters are reliably determined. This is especially useful when the plain Euclidean distance is insufficient for bursts whose reported uncertainties are relatively large. 

The objective is to determine the ratio
\begin{align}\label{eq:metric_output} \frac{d}{\sigma_d}, \end{align}
where $d$ is the Euclidean distance and $\sigma_d$ is its standard uncertainty, derived using the variance formula for error propagation \citep{ku1966}. This ratio is used as the distance metric for DBSCAN, taking measurement uncertainties into account.

In order to derive $\sigma_d$, we begin by computing the partial derivatives of the Cartesian coordinates of the two bursts ($x_i, y_i, z_i$), with respect to $\alpha$, $\delta$ and DM.
By applying the variance formula, the standard uncertainties of the coordinates become:
\begin{align*}
    \sigma_{x_i} = \sqrt{\bigg(\frac{\partial x_i}{\partial \alpha_i}\bigg)^2 \sigma_{\alpha_i}^2 + \bigg(\frac{\partial x_i}{\partial \delta_i}\bigg)^2 \sigma_{\delta_i}^2 + \bigg(\frac{\partial x_i}{\partial \mathrm{DM}_i}\bigg)^2 \sigma_{\mathrm{DM}_i}^2} \\
    \sigma_{y_i} = \sqrt{\bigg(\frac{\partial y_i}{\partial \alpha_i}\bigg)^2 \sigma_{\alpha_i}^2 + \bigg(\frac{\partial y_i}{\partial \delta_i}\bigg)^2 \sigma_{\delta_i}^2 + \bigg(\frac{\partial y_i}{\partial \mathrm{DM}_i}\bigg)^2 \sigma_{\mathrm{DM}_i}^2} \\
    \sigma_{z_i} = \sqrt{\bigg(\frac{\partial z_i}{\partial \alpha_i}\bigg)^2 \sigma_{\alpha_i}^2 + \bigg(\frac{\partial z_i}{\partial \delta_i}\bigg)^2 \sigma_{\delta_i}^2 + \bigg(\frac{\partial z_i}{\partial \mathrm{DM}_i}\bigg)^2 \sigma_{\mathrm{DM}_i}^2},
\end{align*}\\
where $\sigma_{\alpha_i}$, $\sigma_{\delta_i}$ and $\sigma_{\mathrm{DM}_i}$ are the reported errors of the right ascension, declination and DM, respectively.

Because repeaters have been observed to exhibit DM variations up to an order of 3\% \citep{dai2022}, $\sigma_{\mathrm{DM}_i}$ is determined by the maximum value between the measurement uncertainty and the largest expected DM variation.
Thus, the final standard uncertainty becomes:
\begin{align}
    \sigma_d &= \sqrt{\sum_{u \in \{x,y,z\}} \bigg(\frac{\partial d}{\partial u_1}\bigg)^2 \sigma_{u_1}^2+\bigg(\frac{\partial d}{\partial u_2}\bigg)^2 \sigma_{u_2}^2}.
\end{align}
\renewcommand{\arraystretch}{1.2}
\begin{table}
	\centering
	\caption{Fundamental burst properties available for direct plotting through the platform, along with a brief description for each parameter.}
	\label{tab:plot_parameters}
	\begin{tabular}{lrrr} 
		\hline
		\hline
		\textbf{Parameter} & \textbf{Description}\\
		\hline
		\multicolumn{2}{c}{Source Position} \\
		\hline
		\texttt{RA} & \text{Right ascension (HMS)}\\
  		\texttt{RA Error} & \text{Right ascension error ($\arcmin$)}\\
		\texttt{Dec.} & \text{Declination (DMS)}\\
    	\texttt{Dec. Error} & \text{Declination error ($\arcmin$)}\\
		\texttt{Gal. Long.} & \text{Galactic longitude ($l$)}\\
		\texttt{Gal. Lat.} & \text{Galactic latitude ($b$)}\\
		\hline
	    \multicolumn{2}{c}{Burst Properties} \\
		\hline
		\texttt{Frequency} & \text{Reported burst frequency (MHz)}\\
		\texttt{DM} & \text{Reported dispersion measure $\big(\mathrm{pc\ cm}^{-3}\big)$}\\
  		\texttt{DM Error} & \text{Dispersion measure error $\big(\mathrm{pc\ cm}^{-3}\big)$}\\
		\texttt{Flux Density} & \text{Peak flux density (Jy)}\\
		\texttt{Burst Width} & \text{Derived burst width (ms)$^a$}\\
		\texttt{Fluence} & \text{Measured burst fluence (Jy ms)}\\
		\hline
	\end{tabular}
 \\ \  \\ $\ \ ^a$ Commonly reported as the fitted full width at half maximum (FWHM) of the burst, generally referring to the measured width (prior to scattering correction), depending on the submitted entry.

\end{table}
Assuming the threshold for the number of neighbours, \texttt{minPts}, is equal to $2$ (definition of FRB repeater; i.e., present at least 1 detection following the primary burst), we find that, by using Equation~(\ref{eq:metric_output}) as a metric for the distance function, reliable results are obtained when the radius\footnote{For the exact definition of $\epsilon$, see: \url{https://scikit-learn.org/stable/modules/generated/sklearn.cluster.DBSCAN}} $\epsilon$ \citep[arbitrary distance measure;][]{schubert2017} is within a certain range. Given the condition
\begin{align}
    \epsilon \geq \frac{d}{\sigma_d}
\end{align}
employed by DBSCAN to evaluate the association between two bursts, the radius value was optimized (fine-tuned) by identifying the $\epsilon$ threshold that minimizes the number of false positives (FP).

In order to evaluate the confusion matrix and benchmark the robustness of the clustering algorithm, the TNS database was used as a reference for the set of actual (true) repeater parents (22 FRBs).

While the optimal value of $\epsilon \approx 1.15$ was identified using brute-force for simplicity, we note that the global minimum is easily retrievable in a matter of minutes (instead of hours), using conventional optimization algorithms. This is because the problem is one-dimensional and the objective function does not present many instabilities.

By using the \texttt{cluster.DBSCAN} class provided by the \texttt{scikit-learn} library \citep{scikit-learn}, we identify 23 repeating FRB sources. All repeaters reported on the TNS have been successfully identified, yielding a perfect false negative rate:

\begin{equation}
    \mathrm{FNR}={\frac{\mathrm{FN}}{\mathrm{FN}+\mathrm{TP}}}=0,
	\label{eq:fnr}
\end{equation}
where $\mathrm{FN}$ and $\mathrm{TP}$ are the false negatives and true positives, respectively.

The clustering algorithm identified one additional candidate repeater source: FRB 20190425B with one additional sub-burst association (FRB 20190614A). Upon visual inspection, we find that this candidate parent/daughter association is located near the north celestial pole, and has a similar DM with a variation of $\sim$3\%, consistent with what has previously been observed for FRB 20190520B \citep{dai2022} and FRB~121102 \citep{spitler2016, law2017}. We conclude that based on the available information, this source (highlighted in red on the FRBSTATS platform) could be a true repeater that requires further investigation.

Given the efficiency of the algorithm, repeater clustering of 818 bursts is carried out in $\sim$60 seconds by the FRBSTATS server (1 Intel vCPU; 1 GB of memory). This includes the total data read time of the catalogue, the clustering itself, as well as the plotting of the results. The output of the clustering algorithm for 818 events is provided in Fig.~\ref{fig:repeaters}.\\

\subsubsection{Repeater tree visualization}

Once repeater classification is complete, the data is stored as a JSON file, which is finally picked up by the frontend and displayed to the user in the form of a repeater tree for convenient visualization. Upon clicking on an FRB repeater source, its time series plot is displayed, along with the corresponding telescope and signal-to-noise ratio of each detected sub-burst.
\vfill\eject

\subsection{Plots}
\label{sec:plots}

Since the analysis of burst properties is a crucial step required to derive parameter distributions of FRBs and conduct population studies, the platform also offers an easy-to-use plotting tool for quick analysis. Similar to the Australia Telescope National Facility (ATNF) Pulsar Catalogue \citep{manchester2005}, FRBSTATS enables direct parameter visualizations, without the need to download the catalogue and manually run scripts locally, in an offline manner. The burst parameters available for plotting are listed in Table~\ref{tab:plot_parameters}.

\begin{figure}
	\includegraphics[width=1.1\columnwidth]{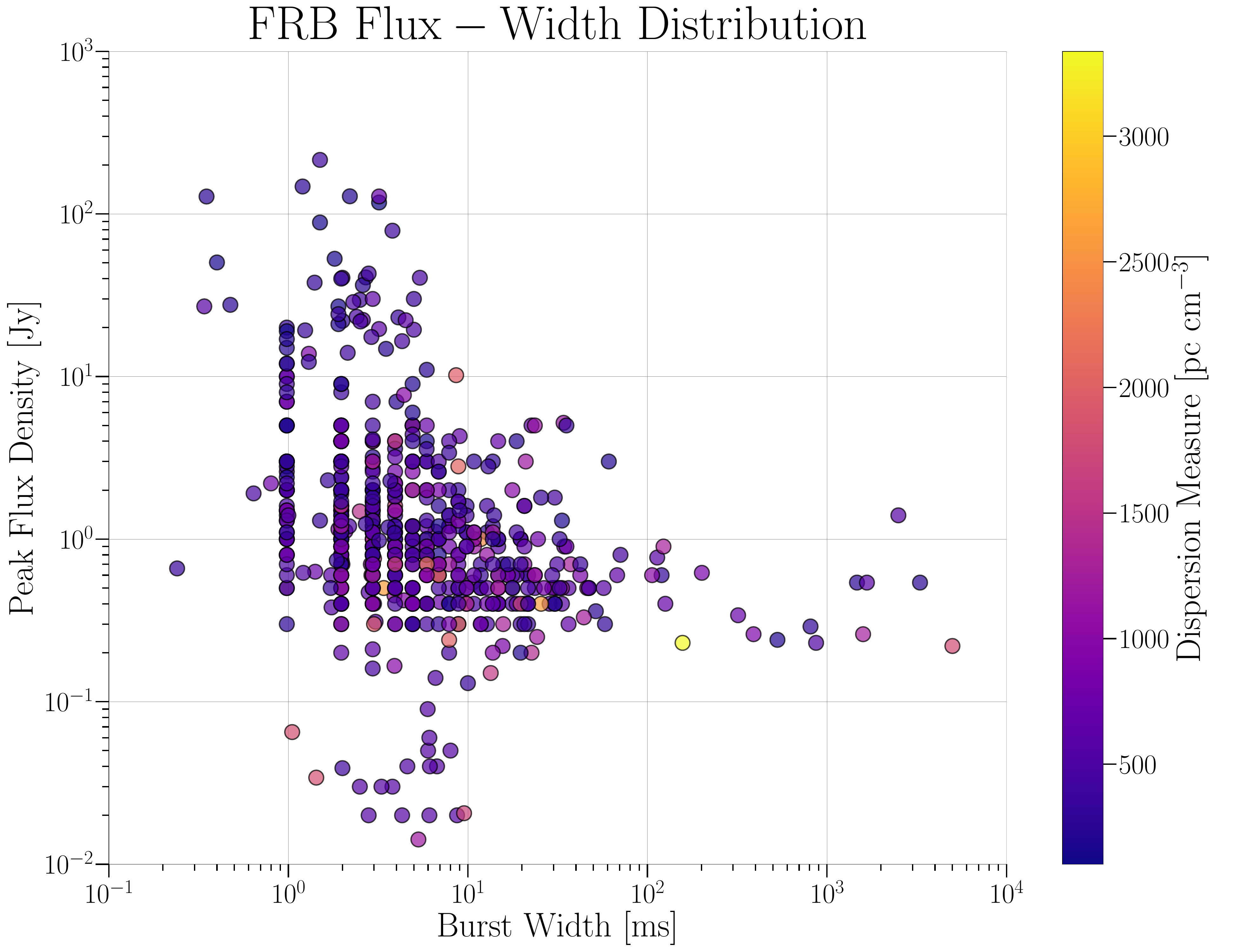}
    \caption{Flux--width distribution of all reported bursts, along with their dispersion measure. Bursts whose flux or width have not been measured and published have been excluded from the chart.\\}
    \label{fig:example_plot}
\end{figure}

An example plot obtained using FRBSTATS is shown in Fig.~\ref{fig:example_plot}, highlighting the distribution of the flux, width, and dispersion measure of bursts. These plots are maintained up to date automatically, and are updated as soon as a new database entry/modification is submitted to the system. This ensures that newly detected events are displayed, and simple comparisons with the remaining FRB population can be made fairly easily.

\subsection{API}
\label{sec:api}

An application programming interface (API) enables a convenient integration layer between a server and its clients. In the case of this particular platform, the FRBSTATS API provides researchers and developers access to FRB parameters in a lightweight, rapid, and easy manner, supporting numerous automation applications. All entries present in the database can be obtained with the help of the API, whose functionality is documented on the web platform, along with some examples. The client may submit a GET request to the provided endpoint URL, and query a burst of interest. If the entry is found in the database, the JSON-encoded API response will consist of the parameters of the queried burst, as well as the reference linked to the corresponding catalogue/publication. Thus, each parameter value can be reliably traced back to its original publication source for further verification. In the case of queries that are not associated with any burst, the API returns an empty array (i.e.,~\texttt{[]}).

It is worth noting that the PHP-based utility automatically neglects the \texttt{`FRB'} prefix, underscores and spaces, in order to improve the parsing of the query and handle different designations (corresponding to the same event) in a compatible fashion. Finally, unlike the catalogue-querying system in place, the server ensures that the search queries submitted to the API are directed to solely match the name of the burst of interest, and does not return irrelevant entries that could happen to match the values of certain parameters.

\section{Conclusions}
\label{sec:conclusions}

We have built and presented a versatile web platform consisting of an accurate open-access catalogue of fast radio bursts, a parameter visualization utility, as well as an application programming interface available for querying burst properties, such as inferred/host redshifts. Furthermore, we expanded on the method of automatically classifying FRBs into repeaters and non-repeaters using the DBSCAN algorithm with a custom distance metric. We hope that the platform provides researchers with access to an up-to-date database of FRBs to work with and conduct better population studies, with as little maintenance work as possible.

In future work, we plan on further expanding the catalogue by providing additional information regarding each burst, such as the Galactic, intergalactic, and local contributions to the DM, as well as polarization properties. Additionally, we hope to make dynamic spectra of each individual burst available and accessible, directly through the online catalogue.

\section*{Acknowledgements}

We wish to acknowledge the help provided by Dr.~John~Antoniadis for the valuable feedback collected and provided to improve the utility of the platform, and Yannick De Clercq for assisting with the deployment of the system.

\bibliographystyle{mnras}
\bibliography{frbstats.bib}

\end{document}